\documentstyle[12pt]{article}
\newcommand{\be}[1]{\begin{equation} \label{(#1)}}
\newcommand{\ee}{\end{equation}}
\newcommand{\ba}[1]{\begin{eqnarray} \label{(#1)}}
\newcommand{\ea}{\end{eqnarray}}
\newcommand{\nn}{\nonumber}
\newcommand{\rf}[1]{(\ref{(#1)})}

\def\pmb#1{\setbox0=\hbox{#1}%
  \kern-.015em\copy0\kern-\wd0
  \kern.03em\copy0\kern-\wd0
  \kern-.015em\raise.0233em\box0 }
\def \znbb {0\nu\beta\beta}

\begin{document}

\begin{center}

{\bf New Leptoquark Mechanism of Neutrinoless Double Beta Decay}

\bigskip

{M. Hirsch, H.V. Klapdor-Kleingrothaus and S.G. Kovalenko$^*$
\bigskip

{\it
Max-Planck-Institut f\"{u}r Kernphysik, P.O. 10 39 80, D-69029,
Heidelberg, Germany}

\bigskip

$^*${\it Joint Institute for Nuclear Research, Dubna, Russia}
}
\end{center}

\begin{abstract}
A new mechanism for neutrinoless double beta ($\znbb$) decay based 
on leptoquark exchange is discussed. Due to the specific helicity 
structure of the effective four-fermion interaction this contribution 
is strongly enhanced 
compared to the well-known mass mechanism of $\znbb$ decay. 
As a result the corresponding leptoquark parameters are severely
constrained from non-observation of $\znbb$-decay. These constraints 
are more stringent than those derived from other experiments. 
\end{abstract}

\noindent
{\it PACs:} 11.30, 12.30, 13.15, 14.80, 23.40


Neutrinoless double beta decay ($\znbb$) is forbidden in the standard model
(SM) of electro-weak interactions since it violates lepton number
conservation. Therefore, experimental observation of this
exotic process would be an unambiguous signal of physics beyond the SM
(see refs. \cite{hax84}-\cite{Trento} for reviews).

Essential progress in the exploration of $\znbb$-decay
both from theoretical and experimental sides has been achieved in the 
last few years (see, for instance \cite{Trento} and references therein).
The considerably improved experimental lower bounds on the half lives 
of various isotopes enhance the potential of $\znbb$ experiments in 
testing different concepts of physics beyond the SM such as
supersymmetry (SUSY) and leptoquarks (LQ).

The SUSY mechanisms of $\znbb$-decay were comprehensively investigated
in a series of papers \cite{Mohapatra}-\cite{HKK2}. It turned out 
that constraints on certain SUSY-parameters from non-observation of 
$\znbb$-decay \cite{HKK1} are stronger than those from current 
and near future accelerator and non-accelerator experiments. 
Therefore, it is useful to investigate other possible 
contributions of physics beyond the SM to $\znbb$-decay to obtain 
$\znbb$ constraints on the corresponding parameters.

In this note we present a new mechanism of $\znbb$-decay associated
with the leptoquark contribution to the effective low-energy
charged current lepton-quark interactions.  The diagrams describing
this contribution are presented in fig. 1.

The SM symmetries allow 5 scalar ($S$) and 5 vector ($V^{\mu}$) LQs
with the following
$LQ(SU(3)_c\otimes SU(2)_L\otimes U(1)_Y)$ assignments:
$S_0(3_c, 1; -2/3)$, $\tilde S_0(3_c, 1; -8/3)$, 
$S_{1/2}(\bar 3_c, 2; -7/3)$,
$\tilde S_{1/2}(\bar 3_c, 2; -1/3)$, $S_1(3_c, 3; -2/3)$,
$V_0(\bar 3_c, 1; -4/3)$,
$\tilde V_0(\bar 3_c, 1; -10/3)$, $V_{1/2}(3_c, 2; -5/3)$,
$\tilde V_{1/2}(3_c, 2; 1/3)$, $V_1(\bar 3_c, 3; -4/3)$, where
$Y = 2(Q_{\it em} - T_3)$.

The most general form of the renormalizable LQ-quark-lepton interactions
consistent with $SU(3)_c\otimes SU(2)_L\otimes U(1)_Y$ gauge symmetry
can be written as \cite{Lagr1}
\ba{LQ-l-q} \nn
&&{\cal L}_{LQ-l-q} =
\lambda^{(R)}_{S_0}\cdot \overline{u^c_R} e_R \cdot S_0^{R\dagger} +
\lambda^{(R)}_{\tilde S_0}\cdot \overline{d^c_R} e_R \cdot
\tilde{S}_0^{\dagger} +
\lambda^{(R)}_{S_{1/2}}\cdot \overline{u_R} \ell_L
\cdot {S}_{1/2}^{R\dagger} + \\ \nn
&&+\lambda^{(R)}_{\tilde S_{1/2}}\cdot \overline{d_R} \ell_L \cdot
\tilde{S}_{1/2}^{\dagger} +
\lambda^{(L)}_{S_0}\cdot \overline{q^c_L} i \tau_2 \ell_L \cdot
S_0^{L\dagger} +
\lambda^{(L)}_{S_{1/2}}\cdot \overline{q_L}  i \tau_2 e_R \cdot
S_{1/2}^{L\dagger} +\\
&&+ \lambda^{(L)}_{S_1}\cdot \overline{q^c_L}  i \tau_2
\hat{S}_1^{\dagger} \ell_L +
\lambda^{(R)}_{V_0}\cdot \overline{d_R} \gamma^{\mu}  e_R
\cdot V_{0\mu}^{R\dagger} +
\lambda^{(R)}_{\tilde V_0}\cdot \overline{u_R} \gamma^{\mu}  e_R \cdot
\tilde{V}_{0\mu}^{\dagger} + \\ \nn
&&+\lambda^{(R)}_{V_{1/2}}\cdot \overline{d^c_R} \gamma^{\mu}  \ell_L
\cdot {V}_{1/2\mu}^{R\dagger} +
\lambda^{(R)}_{\tilde V_{1/2}}\cdot \overline{u^c_R} \gamma^{\mu}  \ell_L
\cdot
\tilde{V}_{1/2\mu}^{\dagger} +
\lambda^{(L)}_{V_0}\cdot \overline{q_L} \gamma^{\mu}  \ell_L \cdot
V_{0\mu}^{L\dagger} +\\ \nn
&&+\lambda^{(L)}_{V_{1/2}}\cdot \overline{q^c_L} \gamma^{\mu} e_R \cdot
V_{1/2\mu}^{L\dagger} +
\lambda^{(L)}_{V_1}\cdot \overline{q_L} \gamma^{\mu}
\hat{V}_{1\mu}^{\dagger} \ell_L + h.c.
\ea
Here $q$ and $\ell$ are the quark
and the lepton doublets. Following \cite{Lagr1,DBC}
we distinguish $S(V)^{L,R}$ being LQs coupled to the left-handed and
right-handed quarks respectively
(see, however, the discussion on chiral couplings in \cite{LQ}).
For LQ triplets $\Phi_1 = S_1, V_1^{\mu}$ the notation
$\hat{\Phi}_1 = \vec\tau\cdot\vec{\Phi}_1$ is used.

On the same footing the LQ fields couple to the SM Higgs doublet
field $H$. A complete list of the renormalizable LQ-Higgs 
interactions is given in ref. \cite{LQ}. These new interactions
are especially important for $\znbb$-decay, since after electro-weak
symmetry breaking they lead to mixing between different
LQ multiplets. In turn this mixing generates
the effective 4-fermion interactions involving right-handed
leptonic currents. In combination with the ordinary SM  left-handed
charged current interactions the latter produce 
the contribution to $\znbb$-decay shown in the diagrams of fig. 1 
with large enhancement factors.
This type of contribution is absent in the case of decoupled
LQ and Higgs sectors \cite{LQ}.

Under electro-weak symmetry breaking the neutral component of
the SM Higgs field acquires a non-zero vacuum  expectation value, 
$\big<H^0\big>$, 
which creates via LQ-Higgs interaction terms non-diagonal mass
matrices for LQ fields with the same electric charge but from different
$SU(2)_L$ multiplets.
To obtain observable predictions from the LQ-lepton-quark interaction
Lagrangian in eq. \rf{LQ-l-q}, the LQ fields ($I= S, V$) with non-diagonal
mass matrices have to be rotated to the mass eigenstate basis $I'$.
This can be done in the standard way: 
$I(Q) = {\cal N}^{(I)}(Q)\cdot I'(Q)$, 
where ${\cal N}^{(I)}(Q)$ are orthogonal matrices such that
${\cal N}^{(I)T}(Q_I)\cdot {\cal M}_I^2(Q)\cdot {\cal N}^{(I)}(Q)
= Diag\{M_{I_n}^2\}$, with the $M_{I_n}$ being the mass of the relevant
mass eigenstate field $I'$.

Now it is straightforward to derive the effective 4-fermion $\nu-u-d-e$
interaction terms generated by the LQ exchange in the 
upper parts of the diagrams 
in fig. 1. After Fierz rearrangement they take the form  \cite{LQ}
\ba{mix_terms}
{\cal L}_{LQ}^{eff} &=&
(\bar\nu P_R e^c)
\left[\frac{\epsilon_S}{M_S^2} (\bar u P_R d) +
\frac{\epsilon_V}{M_V^2} (\bar u P_L d)\right]- \\ \nn
&-& (\bar\nu \gamma^{\mu} P_L e^c) \\ \nn
&\times &\left[
\left(\frac{\alpha_S^{(R)}}{M_S^2}  + \frac{\alpha_V^{(R)}}{M_V^2} \right)
(\bar u \gamma_{\mu} P_R d) -
\sqrt{2}\left(\frac{\alpha_S^{(L)}}{M_S^2}  
+ \frac{\alpha_V^{(L)}}{M_V^2} \right)
(\bar u \gamma_{\mu} P_L d)\right],
\ea
where
\ba{coeff}
\epsilon_I &=&
2^{-\eta_I}
\left[\lambda_{I_1}^{(L)}\lambda_{\tilde{I}_{1/2}}^{(R)}
\left(\theta_{43}^I(Q_I^{(1)}) 
+ \eta_I \sqrt{2} \theta_{41}^I(Q_I^{(2)})\right) -
\lambda_{I_0}^{(L)}\lambda_{\tilde{I}_{1/2}}^{(R)}
\theta_{13}^I(Q_I^{(1)})\right], \\
\alpha_I^{(L)} &=&
\frac{2}{3 + \eta_I} \lambda_{I_{1/2}}^{(L)}\lambda_{I_1}^{(L)}
\theta_{24}^I(Q_I^{(2)}), \ \ \
\alpha_I^{(R)} = \frac{2}{3 + \eta_I}
\lambda_{I_0}^{(R)} \lambda_{\tilde{I}_{1/2}}^{(R)}
\theta_{23}^I(Q_I^{(1)}).
\ea
$\eta_{S,V}= 1,-1$ for scalar and vector LQs. 
$\theta_{kn}^I(Q)$ is a mixing parameter defined by 
\ba{mp}
\theta_{kn}^I(Q) = \sum_{l} {\cal N}_{kl}^{(I)}(Q) {\cal N}_{nl}^{(I)}(Q)
\left(\frac{M_I}{M_{I_l}(Q)}\right)^2,
\ea
where ${\cal N}^{(I)}(Q)$ are mixing matrix elements for
the scalar $I= S$ and vector $I= V$ LQ fields with electric charges
$Q = -1/3,-2/3$. Common mass scales $M_S$ of scalar
and $M_V$ of vector LQs are introduced for convenience.


Following the well known procedure \cite{doi85} one can find the LQ
contribution to the $\znbb$-decay matrix element for 
the diagrams in fig. 1.
The LQ exchange sectors of these diagrams are described by
the point-like 4-fermion interactions specified by the effective Lagrangian
in eq. \rf{mix_terms}. Their bottom parts are the SM
charged current interactions.  The final formula for the inverse half-life
of $\znbb$-decay reads
\ba{half-life}
T^{-1}_{1/2}(\znbb) = |{\cal M}_{GT}|^2  \frac{2}{G_F^2}\left[
\tilde C_1 a^2 + C_4 b_R^2 + 2 C_5 b_L^2 \right]
\ea
with
\ba{coeff2}
a = \frac{\epsilon_S}{M_S^2} + \frac{\epsilon_V}{M_V^2}, \ \ \
b_{L,R} = \left( \frac{\alpha^{(L,R)}_S}{M_S^2} +
\frac{\alpha^{(L,R)}_V}{M_V^2}\right), \ \
\tilde C_1 = C_1 \left(\frac{{\cal M}_1^{(\nu)}/\left(m_e R\right)}{M_{GT} -
\alpha_2 M_F}\right)^{2}
\ea
In eq. \rf{half-life} the coefficients $C_n$ are defined following  
\cite{doi85}; $m_e$ and $R$ are the electron mass and nuclear radius. 
We kept only the dominant terms in eq. \rf{half-life},  neglecting, 
particularly, terms proportional to the neutrino mass $m_{\nu}$ 
which we assume 
to be very small and put $m_{\nu}=0$ in eq. \rf{half-life}. 
Also mixed terms, such as $a \cdot b_{L/R}$, are not accounted for, 
since these are expected to only slightly affect our numerical 
limits. The new matrix element ${\cal M}_1^{(\nu)}$ was introduced and 
calculated in ref. \cite{HKK2} within the pn-QRPA framework. 
Calculating $C_i$ within the same approach \cite{mut89} 
for the particular case of $^{76}$Ge we have a complete set of nuclear 
structure coefficients in eq. \rf{half-life} (all in units of 
inverse years): 
$|{\cal M}_{GT}|^2 \tilde C_1 = 1.63 \times 10^{-10}$, 
$|{\cal M}_{GT}|^2 C_4 = 1.36 \times 10^{-13}, 
|{\cal M}_{GT}|^2 C_5 = 4.44 \times 10^{-9}$. 

Now we are ready to derive constraints on the LQ parameters
$a, b_{L,R}$ in eq. \rf{half-life}. 
We use the result from the Heidelberg-Moscow $^{76}$Ge experiment 
\cite{hdmo94}
$T_{1/2}^{\znbb}(^{76}Ge, 0^+ \rightarrow 0^+) > 7.4 \times 10^{24}$
$years \  90\% \ c.l.$

Assuming no 
spurious cancellations between the different terms in eq. \rf{half-life} 
we derive the following constraints on the effective LQ 
parameters:
\ba{dbd_constraint}
\epsilon_I \leq 2.4 \times 10^{-9} 
\left(\frac{M_I}{100\mbox{GeV}}\right)^2, \\
\alpha_I^{(L)} \leq 2.3 \times 10^{-10} 
\left(\frac{M_I}{100\mbox{GeV}}\right)^2, \\
\alpha_I^{(R)} \leq 8.3 \times 10^{-8} 
\left(\frac{M_I}{100\mbox{GeV}}\right)^2.
\ea
Recall $I = S, V$.


It is interesting to compare these constraints with the corresponding
constraints from other processes \cite{DBC}.
Consider the helicity-suppressed decay
$\pi\rightarrow e\nu$ which is extremely sensitive to the first two
scalar-pseudoscalar terms in eq. \rf{mix_terms}, leading to
a helicity-unsuppressed amplitude \cite{DBC}. The following constraint
from $\pi\rightarrow e\nu$-decay data was obtained in ref. \cite{LQ}: 
$\epsilon_I \leq 5\times 10^{-7}\left(M_I/100\mbox{GeV}\right)^2 $.
Apparently, the corresponding constraints from $\znbb$-decay
in eq. \rf{dbd_constraint} are more stringent by about two 
orders of magnitude. This confirms that $\znbb$-decay is a powerful 
probe of physics beyond the standard model. 

In summary, non-observation of $\znbb$ decay can provide stringent 
bounds on parameters of extensions of the standard model. Moreover, 
the $\znbb$ decay bounds on some of these fundamental parameters can 
be much more stringent than those from other experiments.
Previously such a conclusion was obtained for the case of the 
R-parity violating supersymmetric contribution to $\znbb$-decay 
\cite{HKK1}-\cite{HKK2}. 
In this letter we have shown that the leptoquark mechanism allows 
similar conclusions.

\bigskip
\centerline{\bf ACKNOWLEDGMENTS}

We thank V.A.~Bednyakov, D.I.~Kazakov for helpful discussions. 
M.H. would like to thank the Deutsche Forschungsgemeinschaft 
for financial support by grants kl 253/8-1 and 446 JAP-113/101/0.

{\large\bf Figure Captions}\\

\begin{itemize}
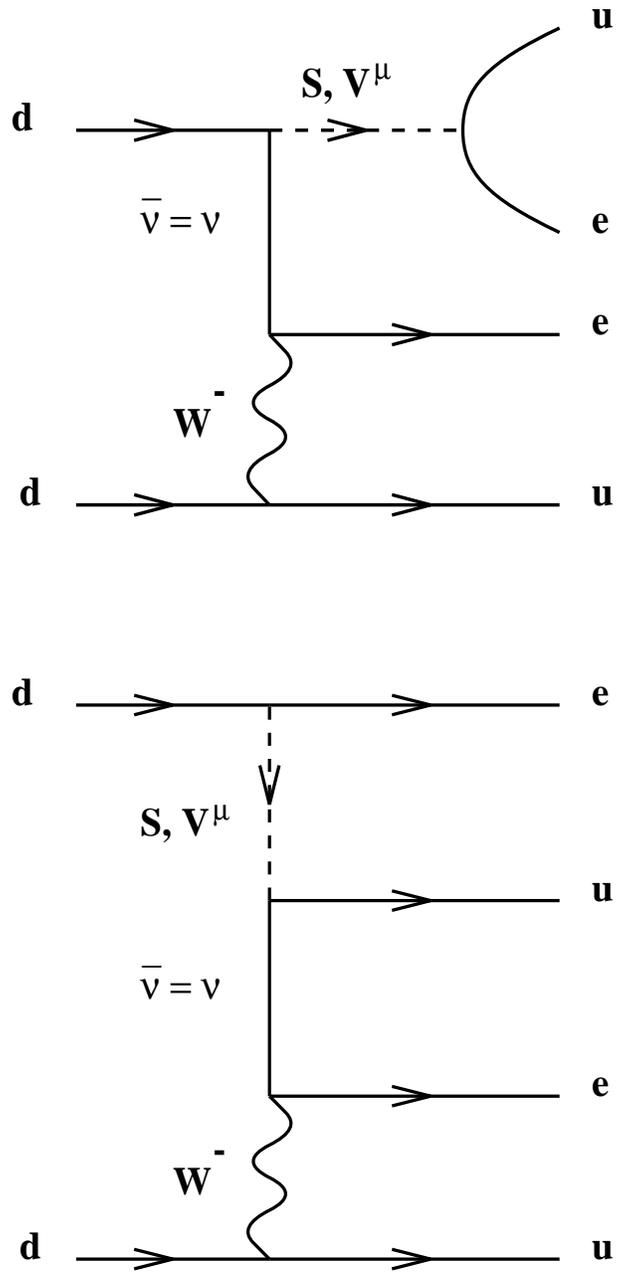
\item[Fig.1]
Feynman graphs for the leptoquark-induced mechanism
of $\znbb$ decay. $S$ and $V^{\mu}$ stand symbolically for 
a) $Q = -1/3$ (upper part) and b) $Q = 2/3$ (lower part) scalar 
and vector LQs. 
\end{itemize}

\end{document}